# Various Carbon to Carbon Bond Lengths Inter-related via the Golden Ratio, and their Linear Dependence on Bond Energies

Raji Heyrovska

Institute of Biophysics, Academy of Sciences of the Czech Republic.

Email: rheyrovs@hotmail.com

**Abstract:**

This work presents the relations between the carbon to carbon bond lengths in the single, double and triple bonds and in graphite, butadiene and benzene. The Golden ratio, which was shown to divide the Bohr radius into two parts pertaining to the charged particles, the electron and proton, and to divide inter-atomic distances into their cationic and anionic radii, also plays a role in the carbon-carbon bonds and in the ionic/polar character of those in graphite, butadiene and benzene. Further, the bond energies of the various CC bonds are shown to vary linearly with the bond lengths.

## 1. Introduction

A general introduction to the importance of carbon can be found in [1]. Carbon-carbon bond lengths are known to vary from about 1.20 to 1.54 Å [2] depending on the chemical nature of their bonding [2, 3]. Considered here are the covalent single, double and triple bonds, the graphite bond, the butadiene bond and the benzene bond (denoted by subscripts: s.b., d.b., t.b., g.b., bt.b. and bz.b. respectively). The bond lengths (+/- 0.01 Å) [2] are given in Figs. 1 a-g. The covalent bonding [2] or the equilibrium [4] radii (R, with the appropriate subscripts) of the carbon atoms are defined [2, 4] as half the respective CC bond lengths. Most of these bond lengths are shown here to be related to each other and through the Golden ratio, $\phi$. The latter (also called The Divine ratio) appears in the geometry of a variety of Nature's creations from microscopic to macroscopic dimensions [5]. The important role of the latter in assigning ionic radii from inter-atomic distances was discovered [6] while interpreting the ground state ionization potential and the Bohr radius of the hydrogen atom. Also, the various radii like the van der Waals radii, valence shell radii and covalent radii of atoms were shown [7] to be linear functions of the respective Bohr radii (obtained from their first ionization potentials) for all the main Group elements in the Periodic Table.

## 2. Bohr radius, atomic and ionic radii and the Golden ratio

The ground state ionization energy ($E_H$) of the hydrogen atom is the energy needed to separate the two oppositely charged particles, the electron ($e^-$) and the proton ($p^+$), from each other to ionize the atom. The ionization potential is thus the difference in the potentials at ionization of the electron and proton given by [6],

$$I_H = E_H/e = (e/2\kappa a_B) = (1/2)[(e/\kappa a_{B,p}) - (e/\kappa a_{B,e})] \tag{1a}$$

$$(1/a_B) = (1/a_{B,p}) - (1/a_{B,e}) \tag{1b}$$

$$(a_{B,e}/a_{B,p})^2 = (a_{B,e}/a_{B,p}) + 1 \tag{1c}$$

$$(a_{B,e}/a_{B,p}) = (1 + 5^{1/2})/2 = 1.618... = \phi \tag{1d}$$

where the Bohr radius $a_B = a_{B,p} + a_{B,e}$ is the sum of the distances pertaining to the proton and electron, respectively, and Eq. 1b is obtained from Eq. 1a. The Golden quadratic Eq. 1c gives the positive root, $(a_{B,e}/a_{B,p}) = \phi$. Thus, $a_B$ has two Golden sections, $a_{B,p} = a_B/\phi^2 = 0.382a_B$ and $a_{B,e} = a_B/\phi = 0.618a_B$ pertaining to $p^+$ and $e^-$ respectively. Also, the <u>ground state spectroscopic energy</u> $E_H$ is itself <u>is a difference between two terms</u>!

This led to the finding [6] that the covalent bond length, d(HH) in the simplest $H_2$ molecule [which was considered by Pauling [2] as consisting of two distances corresponding to the ionic resonance forms, H(+) and H(-), at the same equilibrium distance d(HH)] consists of the Golden ratio based cationic H(+) and anionic H(-) radii of H [6]. It was further shown [6] that the bond length d(AA), in general, between any two atoms (A) of the same kind is the sum of the two Golden sections,

$$d(AA) = 2R(A) = d(AA)/\phi^2 + d(AA)/\phi = R(A+) + R(A-) \qquad (2)$$

where $R(A) = d(AA)/2$ is the covalent or bonding radius, $1/\phi^2 = 0.382$, $1/\phi = 0.618$ and $R(A+) = d(AA)/\phi^2$ and $R(A-) = d(AA)/\phi$ are the Golden ratio based cationic (+) and anionic (-) radii, respectively, of A. This explains the generally known [2] order of the radii, $R(A-) > R(A) > R(A+)$ for elements.

Numerous examples confirming Eq. 2 can be found in [6]. For example, the inter-ionic distances in alkali halides were shown [6] to be exact sums of the Golden ratio based ionic radii (and hence the radius ratio corrections in [2] are unnecessary).

## 3. Single, double and triple bond lengths and their Golden sections

The carbon, carbon single bond length $d(CC)_{s.b.}$ (as in diamond and ethane), double bond length, $d(CC)_{d.b.}$ (e.g., as in ethylene [2]) and triple bond distance $d(CC)_{t.b.}$ (e.g., as in acetylene [2]) have the Golden sections given by,

$$d(CC)_{s.b.} = R(C+)_{s.b} + R(C-)_{s.b} = 0.59 + 0.95 = 2R_{s.b.} = 1.54 \text{ Å} \qquad (3)$$

$$d(CC)_{d.b.} = R(C+)_{d.b} + R(C-)_{d.b} = 0.51 + 0.83 = 2R_{d.b.} = 1.34 \text{ Å} \qquad (4)$$

$$d(CC)_{t.b.} = R(C+)_{t.b} + R(C-)_{t.b} = 0.46 + 0.75 = 2R_{t.b.} = 1.21 \text{ Å} \qquad (5)$$

where $R(C+) = 0.382d(CC)$, $R(C-) = 0.618d(CC)$ and $R = d(CC)/2$ are the respective cationic, anionic and covalent atomic radii.

## 4. Single, double and triple bond radii related via the Golden ratio

These radii, which are half the corresponding $d(CC)$ bond lengths, have the values: $R_{s.b.} = 0.77$ Å, $R_{d.b.} = 0.67$ Å, $R_{t.b.} = 0.605$ Å, (see Figs. 2a,e,f). $R_{s.b.} = 0.77$ Å, is the tetrahedral <u>single bond radius</u> as shown in Fig. 2a. The outer six thin lines are edges with length $AB = a$. (In a regular tetrahedron, the ratio, $R_{s.b.}/a = (3/8)^{1/2} = 0.612$, [8] which is close to $0.618 = 1/\phi$).

The <u>double bond radius</u>, $R_{d.b.} = CF$, the perpendicular distance in Fig. 2e from the center C to the opposite side of an equilateral triangle with $R_{s.b.} = 0.77$ Å as the three equal sides (see the triangle ABC in Fig. 2e). Thus,

$$R_{d.b.} = 0.87 R_{s.b.} = R_{s.b.} \cos 30^0 = (3^{1/2}/2) R_{s.b.} = 0.67 \text{ Å} \qquad (6a)$$

$$R_{s.b.} : R_{d.b.} = 1 : 3^{1/2}/2 = 1 : 0.87 \qquad (6b)$$

where $\cos 30^0 = (3^{1/2}/2)$. In Fig. 2e, in the right angled triangle ACF, $AF = R_{s.b.} \sin 30^0 = R_{s.b.}/2$. [Note: from Eq. 1d that $1/2 = \phi - (5^{1/2}/2)$].

The <u>triple bond radius</u>, $R_{t.b.}$, is related to $R_{s.b.}$ and $R_{d.b.}$ via the Golden ratio thus:

$$R_{t.b.} = 0.786\, R_{s.b.} = R_{s.b.}/\phi^{1/2} = (2/3^{1/2}) R_{d.b.}/\phi^{1/2} = 0.908\, R_{d.b.} = 0.61 \text{ Å} \qquad (7a)$$

$$R_{s.b.} : R_{t.b.} = 1 : \phi^{-1/2} = 1 : 0.79 \qquad (7b)$$

$$R_{d.b.} : R_{t.b.} = 1 : 2\phi^{-1/2}/3^{1/2} = 1 : 0.91 \qquad (7c)$$

where $0.786 = \phi^{-1/2} = \cos 38.17^0$. Thus, in the right angled triangle ACO in Fig. 2f, $R_{s.b.} = AC = 0.77$ Å, $R_{t.b.} = CO = R_{s.b.}\, \phi^{-1/2} = 0.61$ Å, the height from the center (C) of the tetrahedron (not regular) to the center (O) of the base, the angle ACO = $38.17^0$, and AO = $R_{s.b.} \sin 38.17^0 = R_{s.b.}/\phi = 0.476$ Å = $R(C-)_{s.b}/2$ (see Eq. 3).

The <u>triple bond radius</u> is also related to the Bohr radius ($a_B$) [= $a_B(-) + a_B(+)$, the sum of the Golden sections] = 0.64 Å for carbon [7]), thus:

$$R_{t.b.} = 0.95 a_B = a_B \cos 18^0 = 0.61 \text{ Å} \qquad (8a)$$

$$R_{t.b.} : a_B = 1 : 1.05 \qquad (8b)$$

where $\sin 18^0 = 1/2\phi = 0.309$. Note that $\cos 36^0 = 0.809 = \phi/2$. In a right angled triangle with $a_B$ as the hypotenuse and $a_B/2\phi = a_B(-)/2$ as one side, $R_{t.b.}$ forms the third side.

## 5. The carbon, carbon bond length in graphite

The <u>CC bond length in graphite</u>, $d(CC)_{g.b.} = 2R_{g.b.} = 1.42$ Å, has the Golden sections,

$$d(CC)_{g.b.} = R(C+)_{g.b} + R(C-)_{g.b} = 0.54 + 0.88 = 2R_{g.b.} = 1.42 \text{ Å} \qquad (9)$$

where $R(C+)_{g.b}$, $R(C-)_{g.b}$ and $R_{g.b.}$ are the cationic, anionic and covalent graphitic atomic radii respectively. Note that $d(CC)_{g.b.}$ is also the (hybrid) sum of the single bond cationic radius, $R(C+)_{s.b}$ and the double bond anionic radius $R(C-)_{d.b}$ (see Eqs. 3 and 4) (as shown in Fig. 3a):

$$d(CC)_{g.b.} = R(C+)_{s.b} + R(C-)_{d.b} = 0.59 + 0.83 = 2R_{g.b.} = 1.42 \text{ Å} \qquad (10)$$

This shows that the <u>CC bond in graphite</u> can also be considered as an ionic bond, $C(+)_{s.b}C(-)_{d.b}$ and is probably the reason for its electrical conduction. The radius $R_{g.b.}$ is also related to $R_{s.b.}$ via the Golden ratio as shown:

$$R_{g.b.} = 0.922\, R_{s.b.} = R_{s.b.}(1 - 1/\phi^4)^{1/2} = R_{s.b.}\cos 22.46^0 = 0.71 \text{ Å} \qquad (11)$$

In the triangle ACO in Fig. 2b, $R_{s.b.} = AC$, $R_{g.b.} = CO = R_{s.b.}(1 - 1/\phi^4)^{1/2}$, the angle ACO = $22.46^0$ and $AO = R_{s.b.}\sin 22.46^0 = R_{s.b.}/\phi^2 = R(C+)_{s.b.}/2 = 0.29$ Å (see Eq. 3).

## 6. The carbon, carbon bond length in butadiene

There are two CC bond lengths in butadiene: a double bond (see Eq. 4) and a butadiene bond with $d(CC)_{bt.b.} = 2R_{bt.b.} = 1.46$ Å. The Golden sections are given by,

$$d(CC)_{bt.b.} = R(C+)_{bt.b} + R(C-)_{bt.b} = 0.56 + 0.90 = 2R_{bt.b.} = 1.46 \text{ Å} \qquad (12)$$

Note that $d(CC)_{bt.b.}$ is also the (hybrid) sum as in graphite, but of the second possible combination of the Golden sections of the single and double bonds (Eqs. 3 and 4), (see Fig. 3b), giving it an ionic/polar character as in graphite:

$$d(CC)_{bt.b.} = R(C+)_{d.b} + R(C-)_{s.b} = 0.51 + 0.95 = 2R_{bt.b.} = 1.46 \text{ Å} \qquad (13)$$

Thus, the CC bond in butadiene can also be considered as an ionic bond, $C(-)_{s.b}C(+)_{d.b}$ as in graphite.

The radius $R_{bt.b.}$ is related to $R_{s.b.}$ via the Golden ratio as shown:

$$R_{bt.b.} = 0.948\, R_{s.b.} = R_{s.b.}(1 - 1/4\phi^2)^{1/2} = R_{s.b.}\cos 18^0 = 0.73 \text{ Å} \qquad (14)$$

In the triangle ACO in Fig. 2c, $R_{s.b.} = AC$, $R_{bt.b.} = CO$, the angle ACO = $18^0$ and $AO = R_{s.b.}\sin 18^0 = R_{s.b.}/2\phi = 0.24$ Å. (From Eqs. 8a and 14, $R_{t.b.}/a_B = \cos 18^0 = R_{bt.b.}/R_{s.b.}$.)

## 7. The carbon-carbon bond length in benzene

The six <u>benzene bond lengths</u> $d(CC)_{bz.b.}$ are each equal to 1.38 (+/- 0.01) Å [2] and the Golden sections are:

$$d(CC)_{bz.b.} = R(C+)_{bz.b.} + R(C-)_{bz.b} = 0.53 + 0.85 = 2R_{bz.b.} = 1.38 \text{ Å} \qquad (15)$$

Note that $d(CC)_{bz.b.}$ is a hybrid sum of the covalent radii:

$$d(CC)_{bz.b.} = R(C)_{d.b.} + R(C)_{g.b.} = 0.67 + 0.71 = 2R_{bz.b.} = 1.38 \text{ Å} \qquad (16)$$

Thus, in benzene, three of the six carbon atoms of radii $R(C)_{d.b.} = 0.67$ Å alternate with three of radii $R(C)_{g.b.} = 0.71$ Å and $R_{bz.b.} = 0.69$ Å. See Figs. 1g and 4a.

There are <u>two (hybrid) sums of radii</u> which are close to +/- 0.01 Å of 1.38 Å:

$$d(CC')_{bz.b.} = R(C+)_{d.b.} + R(C-)_{g.b.} = 0.51 + 0.88 = 1.39 \text{ Å} \qquad (17a)$$
$$d(CC'')_{bz.b.} = R(C+)_{g.b} + R(C-)_{d.b} = 0.54 + 0.83 = 1.37 \text{ Å} \qquad (17b)$$

These are shown in Figs. 4c,e. Eqs. 17a,b show that $(CC')_{bz.b.} = C(+)_{d.b.}C(-)_{g.b.}$ and $(CC'')_{bz.b.} = C(-)_{d.b.}(C+)_{g.b.}$ are the ionic forms and hence have a polar/ionic/pi-bond/ character, see Figs 4 c,e.

The radii, $R_{bz.b.} = d(CC)_{bz.b.}/2 = 0.69$ Å and $R_{s.b.}$ are also related via the Golden ratio:

$$R_{bz.b.} = 0.90 \, R_{s.b.} = 2R_{s.b.}/(2\phi - 1) = R_{s.b.}\cos 26.57^0 = 0.69 \text{ Å} \qquad (18)$$

In the triangle ACO in Fig. 2d, $R_{s.b.} = AC$, $R_{bz.b.} = CO$, the angle $ACO = 26.57^0$, $AO = R_{s.b.} \sin 26.57^0 = R_{bz.b.} \tan 26.57^0 = R_{bz.b.}/2 = 0.345$ Å and $CO/AO = 2$.

## 8. Linear dependence of CC bond energy on bond lengths

The bond energy data in [2, 3, 4] for the various CC bond lengths ranging from 1.21 Å (for the triple bond) till 1.54 Å (for the single bond) and the data for benzene from [9] have been assembled in Table 1. Fig. 5 shows the linear dependence of the bond energy on the bond lengths.

**Acknowledgements:** The author thanks Prof. E. Palecek of the IBP, Academy of Sciences of the Czech Republic for the moral support and the IBP for financial support.

**Figs. 1a-g**: The CC bond lengths [2], d(CC):  s.b., bt.b., g.b., bz.b., d.b., t.b. and bz.b.

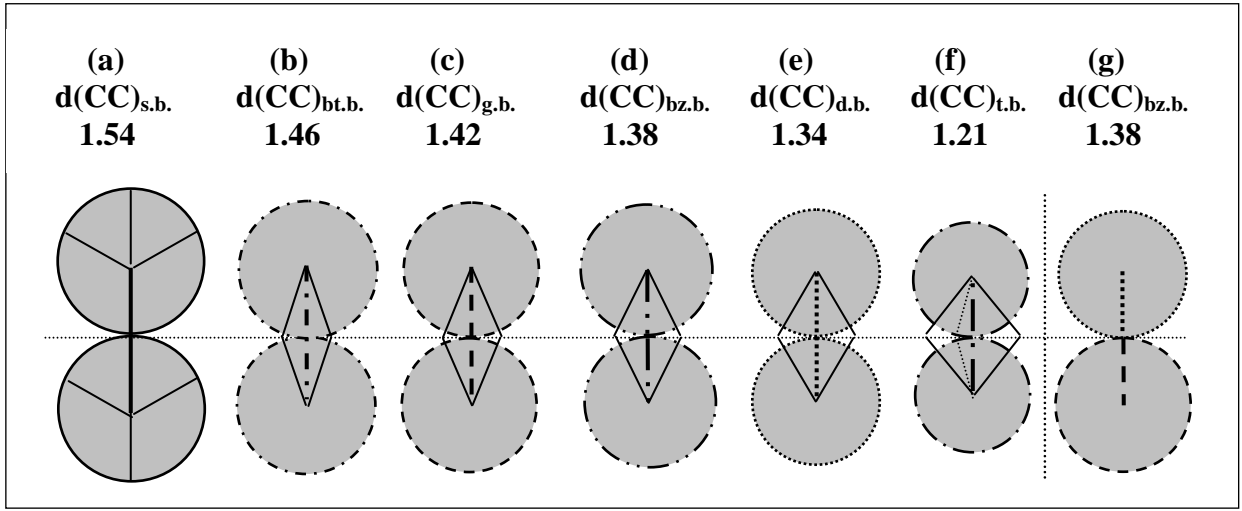

**Figs. 2a-f:** Comparison of all the radii with $R_{s.b.} = 0.77 = CA$ [= CB = CD = CE in (a)] in (a)-(f) (as in Fig. 1a-f). The radii: (e) $R_{d.b.} = 0.67$, (f) $R_{t.b.} = 0.61$, (b), $R_{g.b.} = 0.71$, (c) $R_{bt.b.} = 0.73$ and (d) $R_{bz.b.} = 0.69$ Å. In (e) CA: CF: AF = 1: $3^{1/2}/2$: 0.5 = 1: 0.87: 0.5. The ratio CA: CO: AO: (f) 1: $\phi^{-1/2}$: $1/\phi$ = 1: 0.77: 0.62, (b) 1: $(1 - 1/\phi^4)^{1/2}$: $\phi^{-2}$ = 1: 0.92: 0.38, (c) 1: $(1 - 1/4\phi^2)^{1/2}$: $1/2\phi$ = 1: 0.95: 0.31 and (d) 1: $2/(2\phi - 1)$: $1/(2\phi - 1)$ = 1: 0.89: 0.45.

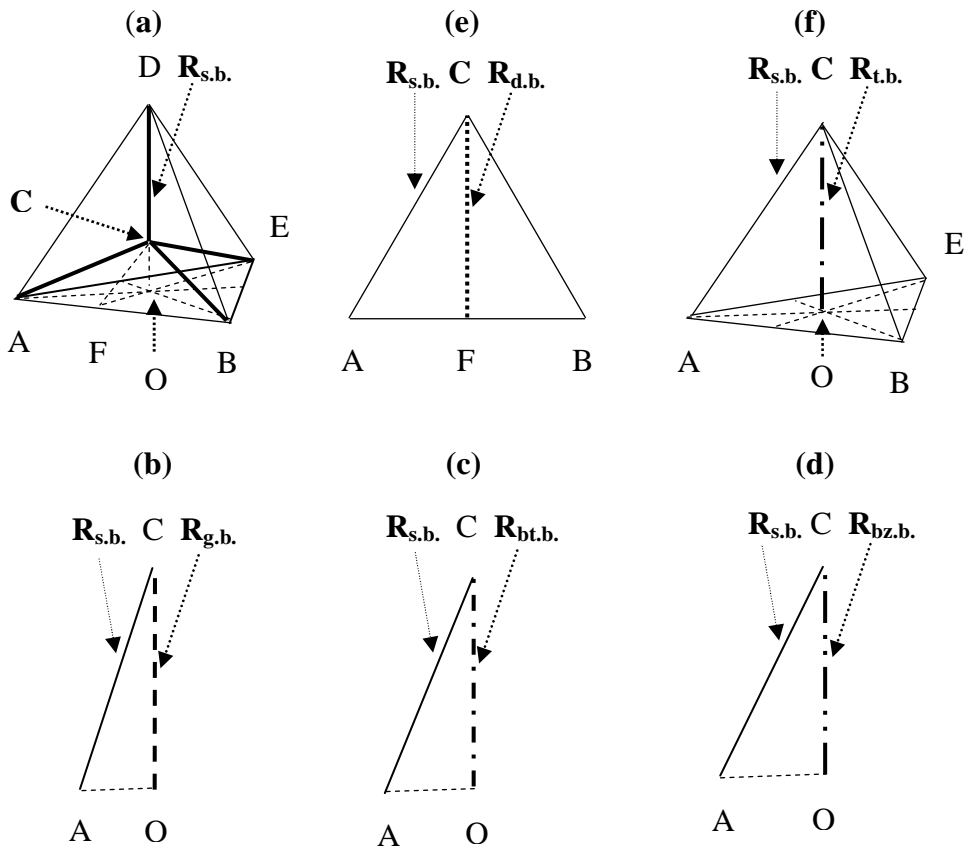

**Fig. 3. (a)** Graphite CC bond length as the hybrid sum, $R_{s.b.}(+) + R_{d.b.}(-) = 0.59 + 0.83 = d(C_{s+}C_{d-}) = d(CC)_{g.b.} = 1.42$ Å. **(b)** Butadiene CC bond length as the hybrid sum, $R_{s.b.}(-) + R_{d.b.}(+) = 0.95 + 0.51 = d(C_{s-}C_{d+}) = d(CC_{bt.b.}) = 1.46$ Å. Arrows: inner circles: cations (+) and outer circles: anions (-).

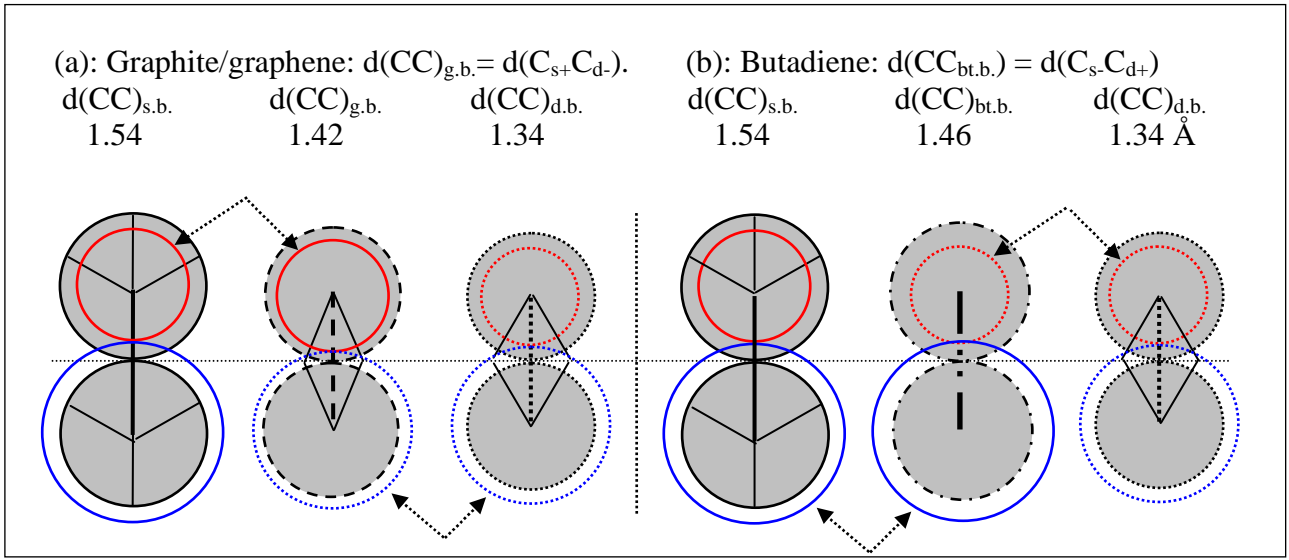

**Fig. 4. (a):** Benzene CC bond length as the covalent radii sum, $d(CC)_{bz.b.} = R_{d.b.} + R_{g.b.} = 0.67 + 0.71 = 2R_{bz.b.} = 1.38$ Å (see also Fig. 1g). The close hybrid sums, **(c):** $d(CC')_{bz.b.} = R(+)_{d.b.} + R(-)_{g.b.} = 0.51 + 0.88 = d(C_g C_{d+}) = 2R'_{bz.b.} = 1.39$ Å and **(e):** $d(CC'')_{bz.b.} = R(+)_{g.b} + R(-)_{d.b} = 0.54 + 0.83 = d(C_{g+}C_{d-}) = 2R''_{bz.b.} = 1.37$ Å. (Arrows: inner circles: cations (+) and outer circles: anions (-)).

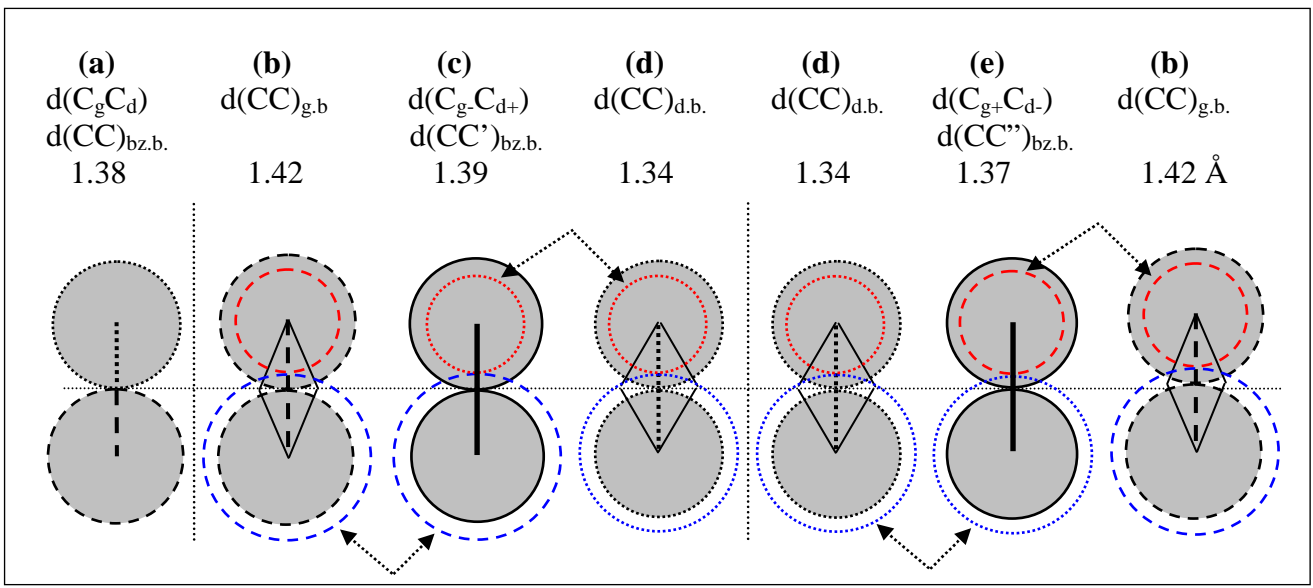

**Fig. 5.** Linear dependence of bond energies on the bond lengths of CC bonds (data in Table 1).

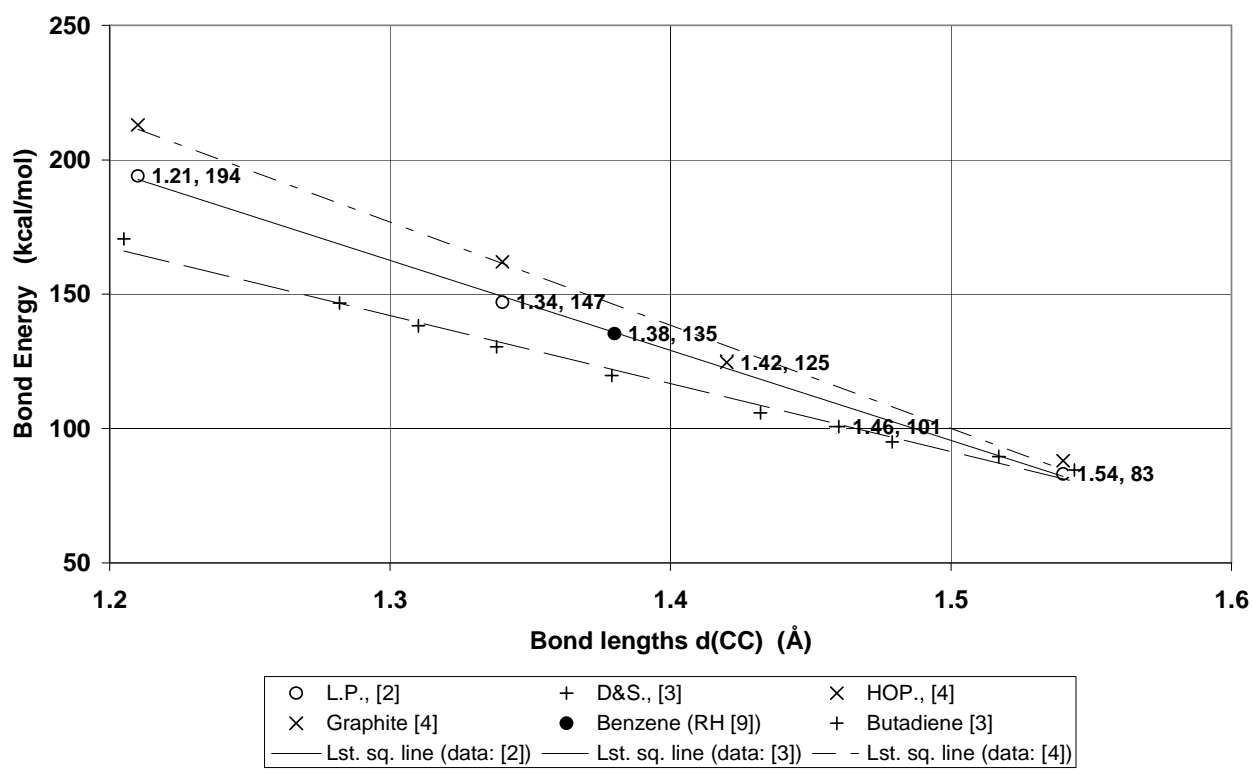

**Table 1.** Carbon-carbon bond lengths and bond energies

| Bonds | d(CC) [2] Å | B.E. kcal/mol | Ref. | B.E. [4] kcal/mol | d(CC) [3] Å | B.E. [3] kcal/mol |
|---|---|---|---|---|---|---|
| s.b | 1.54 | 83 | LP [2] | 88 | 1.54 | 85 |
| d.b | 1.34 | 147 | LP [2] | 162 | 1.52 | 90 |
| t.b | 1.21 | 194 | LP [2] | 213 | 1.46 | 101 |
| g.b | 1.42 | 125 | HOP [4] | 125 | 1.48 | 95 |
| bz.b | 1.38 | 135 | RH [9] |  | 1.43 | 106 |
| bt.b | 1.46 | 101 | D&S [3] |  | 1.38 | 120 |
|  |  |  |  |  | 1.34 | 130 |
|  |  |  |  |  | 1.31 | 138 |
|  |  |  |  |  | 1.28 | 147 |
|  |  |  |  |  | 1.21 | 171 |